\documentclass[aps,prl,amsmath,twocolumn,showpacs,superscriptaddress]{revtex4-1}
\usepackage{amsmath, amssymb, graphicx, bm}
\usepackage[usenames,dvipsnames,svgnames,table]{xcolor}
\usepackage[colorlinks=true,linkcolor=RoyalBlue,citecolor=RoyalBlue]{hyperref}

\begin{document}
\title{Magnonic Spin-Transfer Torque in Ferromagnet/Antiferromagnet/Ferromagnet Trilayer}

\author{Ran Cheng}
\affiliation{Department of Physics, Carnegie Mellon University, Pittsburgh, PA 15213}
\affiliation{Department of Electrical and Computer Engineering, Carnegie Mellon University, Pittsburgh, PA 15213}

\author{Di Xiao} 
\affiliation{Department of Physics, Carnegie Mellon University, Pittsburgh, PA 15213}

\author{Jian-Gang Zhu}
\affiliation{Department of Electrical and Computer Engineering, Carnegie Mellon University, Pittsburgh, PA 15213}
\affiliation{Department of Physics, Carnegie Mellon University, Pittsburgh, PA 15213}

\begin{abstract}

In an antiferromagnet (AF) with uniaxial anisotropy, spin-up and spin-down magnons coexist and form an intrinsic degree of freedom resembling electrons. When polarized by an adjacent ferromagnet (F), a magnonic pure spin current can be thermally generated in an AF. We explore thermal magnon transport in an insulating F/AF/F trilayer where propagating magnons inside the AF spacer can transfer angular momenta between the two Fs. We find that a sufficiently large temperature gradient can switch the downstream F via magnonic spin-transfer torque if it is initially antiparallel with the upstream F. A reciprocal switching is achievable by reversing the temperature gradient. Using typical material parameters, we estimate the threshold to be less than 1 K/nm at room temperature, which can be reduced by raising temperature and enhancing the interfacial exchange coupling.

\end{abstract}

\maketitle

Seeking efficient ways to process and store information with low dissipation has been an everlasting effort in modern electronics. Joining this endeavor is the exploitation of the electron spin degree of freedom. Spin is an intrinsic angular momentum that, unlike charge, can transport information in the form of pure spin currents without incurring Joule heating. In spite of this advantage, generation of a pure spin current often relies on the flow of charge (such as the spin Hall effect~\cite{ref:SHE}), which inevitably creates waste heat.

Magnons, the quanta of spin-wave excitations, can also be utilized to transfer spin, which has garnered significant attention in recent years~\cite{ref:Magnonics,ref:YIG,ref:Bauer,ref:Kajiwara}. Different from electrons, magnons can propagate over long distances in insulating materials without an accompanying charge current, holding the promise that translative motion of electrons are no more needed to transport information. However, in order for magnons to fully function as electrons, the magnon spin should act as an intrinsic degree of freedom besides merely being an angular momentum.

In an antiferromagnet (AF) with easy-axis anisotropy, symmetry ensures that magnon excitations are doubly degenerate at zero magnetic fields~\cite{ref:AFMR}. The two modes are circularly-polarized in opposite fashions and carry opposite spins, forming an internal space capable of encoding information~\cite{ref:AFSP}. Consequently, magnons in an AF can function in many ways as conduction electrons. This unique feature enables us to explore the magnonic counterparts of phenomena usually associated with the electron spin~\cite{ref:FET,ref:SNE1,ref:SNE2,ref:SNEexp,ref:Chirality,ref:SMLocking,ref:MTI}.

The promise of utilizing magnons as active spin carriers has been illuminated in a related scenario: spin transmission across insulating AF~\cite{ref:LWW,ref:HDZ,ref:Kent,ref:Qiu,ref:Zink,ref:Baltz,ref:Ono,ref:Hahn,ref:WHL}. By injecting spins on one side of an AF insulator, one is able to detect survival spins on the other side. However, the observed spin survival rate decays on the order of nanometers and culminates around the ordering temperature, indicating that spin transmission is dominated by either evanescence waves~\cite{ref:Slavin} at low temperatures or spin fluctuations at high temperatures. A propagating magnon picture, however, cannot be corroborated by available evidences. This should not be a surprise since the choice of AFs in existing experiments, \textit{e.g.}\ NiO, does not respect the uniaxial symmetry that guarantees the degeneracy of two circularly-polarized modes. Magnons in these materials are linearly-polarized and non-degenerate~\cite{ref:NiO}, thus losing the analogy with electrons. Recently, the possibility of long-range spin transmission via antiferromagnetic spin dynamics has been proposed~\cite{ref:Duine,ref:Takei}. However, it remains unclear in what phenomena the coexistence of two magnon spin species is a defining factor, and it is an open question whether antiferromagnetic magnons can replace electrons to fulfill tasks that require the presence of opposite spin polarizations.
 
\begin{figure}[t]
	\includegraphics[width=\columnwidth]{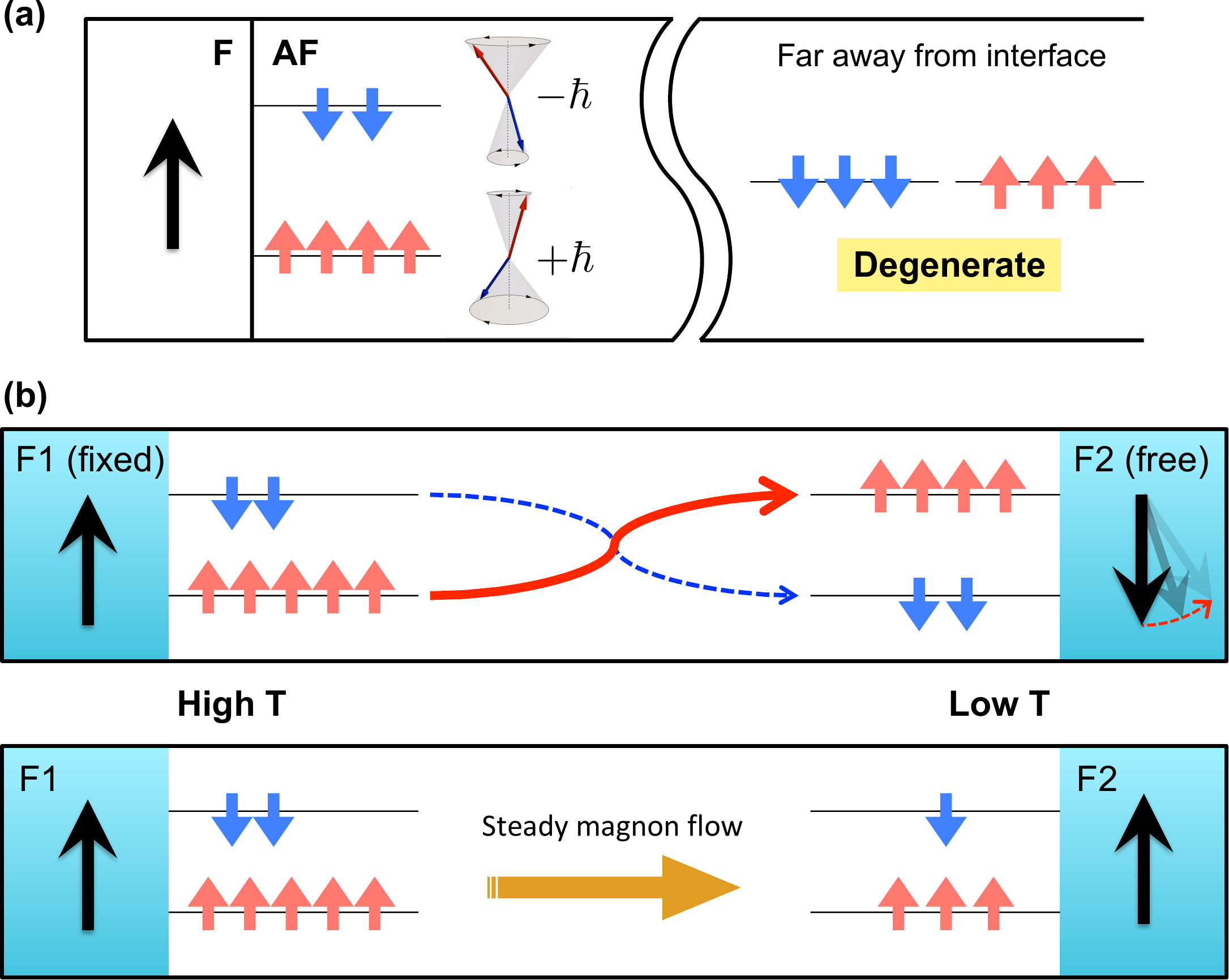}
	\caption{Schematics of the proposed switching mechanism based on a F/AF/F trilayer. (a) Interfacial exchange coupling lifts the degeneracy of the magnons with opposite spins; the degenerate is retrieved away from the interface. (b) In the antiparallel configuration, magnon populations on opposite interfaces are imbalanced in opposite ways. A sufficiently large temperature gradient can deliver majority spins from F1 (hot side) to F2 (cold side), leading to a population inversion that switches the downstream magnetization. A steady magnon spin current is established after the switching process is completed.}
	\label{fig:schematic}
\end{figure}

Basing on the coexistence of both spin species in uniaxial AFs, we propose in this Letter a novel magnonic spin-transfer torque (STT) that can switch an adjacent ferromagnet (F). As schematically illustrated in Fig.~\ref{fig:schematic}, we consider an exchange-coupled F/AF/F trilayer. While the insulating AF blocks charge transport, it behaves as a spin conductor hosting both spin-up and spin-down carriers. Under a temperature gradient $\bm{\nabla}T$, spin angular momenta are delivered from the hot to the cold side, acting on the downstream F. Magnetic switching is triggered when $\bm{\nabla}T$ reaches a threshold that depends on both temperature and the interfacial exchange coupling. A reciprocal switching can be realized by reversing $\bm{\nabla}T$.

\textit{Physical Picture.}---We assume that the two Fs are extremely thin and have a Curie temperature far exceeding the N\'{e}el temperature of the AF. As a result, we ignore spin transport inside the two Fs and simply treat them as two (boundary) spin polarizers for the magnon gas in the AF spacer. As shown in Fig.~\ref{fig:schematic}(a), magnons with opposite spins are degenerate far away from the interface. In the interfacial region, the exchange coupling acts as a local Zeeman field and lifts the degeneracy, leading to a thermal population of spin angular momenta. If the magnetization is collinear with the AF easy-axis, magnons with spins parallel (antiparallel) to the magnetization have a lower (higher) frequency:
\begin{align}
 \omega_{_{\rm P/AP}}=\sqrt{(\Delta/\hbar)^2+v_s^2k^2}\mp J/\hbar, \label{eq:dispersion}
\end{align}
where $\Delta$ is the excitation gap, $\hbar$ is the reduced Planck constant, $v_s$ is the spin-wave velocity, $k$ is the norm of wave vector, and $J=J_0a/L$ is the effective Zeeman field with $L$ the AF thickness, $a$ the lattice constant, and $J_0$ the exchange coupling that connects atoms on both sides of the interface.

If the trilayer is prepared antiparallel, the exchange splitting at different interfaces are opposite. As illustrated in Fig.~\ref{fig:schematic}(b), spin-up (spin-down) magnons are more populated on the left (right) interface. In spite of this difference, no spin current will form since left-moving carriers are balanced by their right-moving partners. We can break this balance by applying a temperature gradient $\nabla T$, which generates a magnon spin current inside the AF. Driven by $\nabla T$, both majority and minority spins (with respect to F1) diffuse from the hot to the cold side, after which they switch roles and deliver angular momenta to F2. When $\nabla T$ is sufficiently large, it will induce a population inversion of magnons on the cold side. By absorbing enough magnons with inverted spins, F2 will switch direction so that the trilayer ends up with a parallel configuration.

To stabilize the switching process without affecting F1, we need to pin F1 by a substrate. This is because angular momentum conservation dictates that F1 must experience an opposite STT during the F2 switching~\cite{ref:STT}. We assume in the following that F1 is a fixed layer whereas F2 is a free layer. When the switching is completed, as shown in Fig.~\ref{fig:schematic}(b), a steady magnon current with spin-up polarization is established. In the parallel configuration, if we reverse the direction of $\bm{\nabla}T$, F2 will be on the hot side and magnons will flow leftward. If an reversed $\nabla T$ is strong enough, F2 will rapidly lose spin-up angular momenta and eventually switch back to the antiparallel state. Therefore, the conjectured switching mechanism is reciprocal with respect to the direction of temperature gradient.

The STT acting on F2 arises from the net spin current flowing into F2. Because of angular momentum conservation, only the transverse spin polarization with respect to F2 contributes~\cite{ref:STT}. Characterizing the instantaneous magnetization of F1 and F2 by two unit vectors $\bm{S}_1$ and $\bm{S}_2$, and labeling the magnon spin polarization with respect to $\bm{S}_1$, we write down the Landau-Lifshitz-Gilbert equation of F2 as
\begin{align}
\frac{\partial\bm{S}_2}{\partial t}=&\frac{K_F}{\hbar}S_2^z\hat{\bm{z}}\times\bm{S}_2+\alpha\bm{S}_2\times\frac{\partial\bm{S}_2}{\partial t} \notag\\
&\qquad+j_s(L)A\bm{S}_2\times(\bm{S}_1\times\bm{S}_2), \label{eq:LLG}
\end{align}
where $\alpha$ is the Gilbert damping constant, $\hat{\bm{z}}$ is the easy-axis, $K_F$ is the anisotropy of F2,  $A$ is the area of the interface, and $j_s(L)$ is the spin current density at $x=L$ scaled in the dimension of a number current density. Now the problem boils down to the calculation of $j_s$ as a function of the temperature gradient.

\textit{Calculation and Results.}---The coordinate system is chosen such that the transport direction is $x$ and the easy-axis is $z$. The two F/AF interfaces are located at $x=0$ and $x=L$, respectively. Similar to the case of electrons~\cite{ref:CIP,ref:Slon}, we assume that magnons only memorize the most recent interfacial collision, losing all prior information. It is the memory loss upon reflection (\textit{i.e.}, loss of spin polarization) that generates reactions on the adjacent F in the form of STTs. In this regard, we use two distribution functions $f^>_{\sigma}(x)$ and $f^<_{\sigma}(x)$ to describe right-moving and left-moving magnons, respectively. As mentioned above, the spin index $\sigma=+(-)$ denotes a spin polarization parallel (antiparallel) to the fixed layer $\bm{S}_1$. Different from electrons in a normal metal, magnons in the AF has a prescribed quantization axis so magnon spins can be either $+\hbar$ or $-\hbar$ with respect to $z$. When F2 deviates from $z$ by an angle $\theta$, the Zeeman field acting on the AF/F2 interface is effectively $J\cos\theta$. Therefore, we can impose the boundary conditions for $f^>_{\sigma}(x)$ and $f^<_{\sigma}(x)$ as: $f_\sigma^>(0)=f_\sigma^0+\sigma J\left(-\frac{\partial f_\sigma^0}{\hbar\partial\omega}\right)$ and $f_\sigma^<(L)=f_\sigma^0+\sigma J\cos\theta\left(-\frac{\partial f_\sigma^0}{\hbar\partial\omega}\right)$, where $f_\sigma^0=1/[e^{\hbar\omega_\sigma/k_BT}-1]$ is the global equilibrium distribution function.

As suggested in the previous section, $f_\sigma^>$ and $f_\sigma^<$ must equal in the absence of $\nabla T$, otherwise a spin current will exist without an external driving force. This balance can be maintained when the system size $L$ is much larger than the momentum relaxation length (\textit{viz.}, mean free path). Specifically, if $f_\sigma^>$ differs from $f_\sigma^<$, they will relax towards thermal equilibrium according to
\begin{subequations}
	\label{eq:Boltz}
	\begin{align}
	v_x^\sigma\frac{\partial f_\sigma^>}{\partial x}&=-\frac{f_\sigma^>-f_\sigma^<}{\tau_p}-\frac{f_\sigma^>-f_\sigma^0}{\tau_{th}},\ (v_x>0) \\
	v_x^\sigma\frac{\partial f_\sigma^<}{\partial x}&=-\frac{f_\sigma^<-f_\sigma^>}{\tau_p}-\frac{f_\sigma^<-f_\sigma^0}{\tau_{th}},\ (v_x<0)
	\end{align}
\end{subequations}
where $v_x^\sigma=\partial\omega_\sigma/\partial k_x$ is the group velocity of magnons with spin $\sigma$. On the right-hand side, the first term describes momentum relaxation, which includes scattering processes that conserve the magnon number and only restore \textit{local} thermal equilibrium. By contrast, the second term describes scattering processes that do not conserve the magnon number, through which magnons lose angular momenta and relax towards the global equilibrium represented by $f_\sigma^0$. As discussed in existing literature, $\tau_p\ll\tau_{th}$ in ferromagnetic insulators~\cite{ref:Shulei,ref:Wees,ref:Rezende,ref:CG}. Here, the same physics applies to the magnon transport of a given spin species in AF. Note that spin-flip scattering that mixes $\sigma=+$ and $\sigma=-$ magnons is possible only in the presence of magnetic impurities breaking the uniaxial symmetry. In this Letter, however, we ignore spin-flip scattering completely. These assumptions combined yields a magnon transport diffusive in momentum but ballistic in spin, which is analogous to the case of electrons in a spin-valve~\cite{ref:SV}. In the regime of $\tau_p\ll L/v_x\ll\tau_{th}$, the $\tau_{th}$-controlled relaxation drops out and Eq.~\eqref{eq:Boltz} gives $f_\sigma^>\approx f_\sigma^<= \bar{f}_\sigma$, where
\begin{align}
 \bar{f}_\sigma=\frac{1}{\exp\left[ \hbar\omega-\mu_\sigma(x) \right]/k_BT-1}
\end{align}
is the local equilibrium distribution function and
\begin{align}
 \mu_\sigma(x)=\sigma J\left[1+(\cos\theta-1)\frac{x}{L}\right] \label{eq:chemical}
\end{align}
is the local chemical potential of spin species $\sigma$. Since $f_\sigma^>\approx f_\sigma^<$, there is no spin current at $\nabla T=0$.

\begin{figure}[t]
	\centering
	\includegraphics[width=\columnwidth]{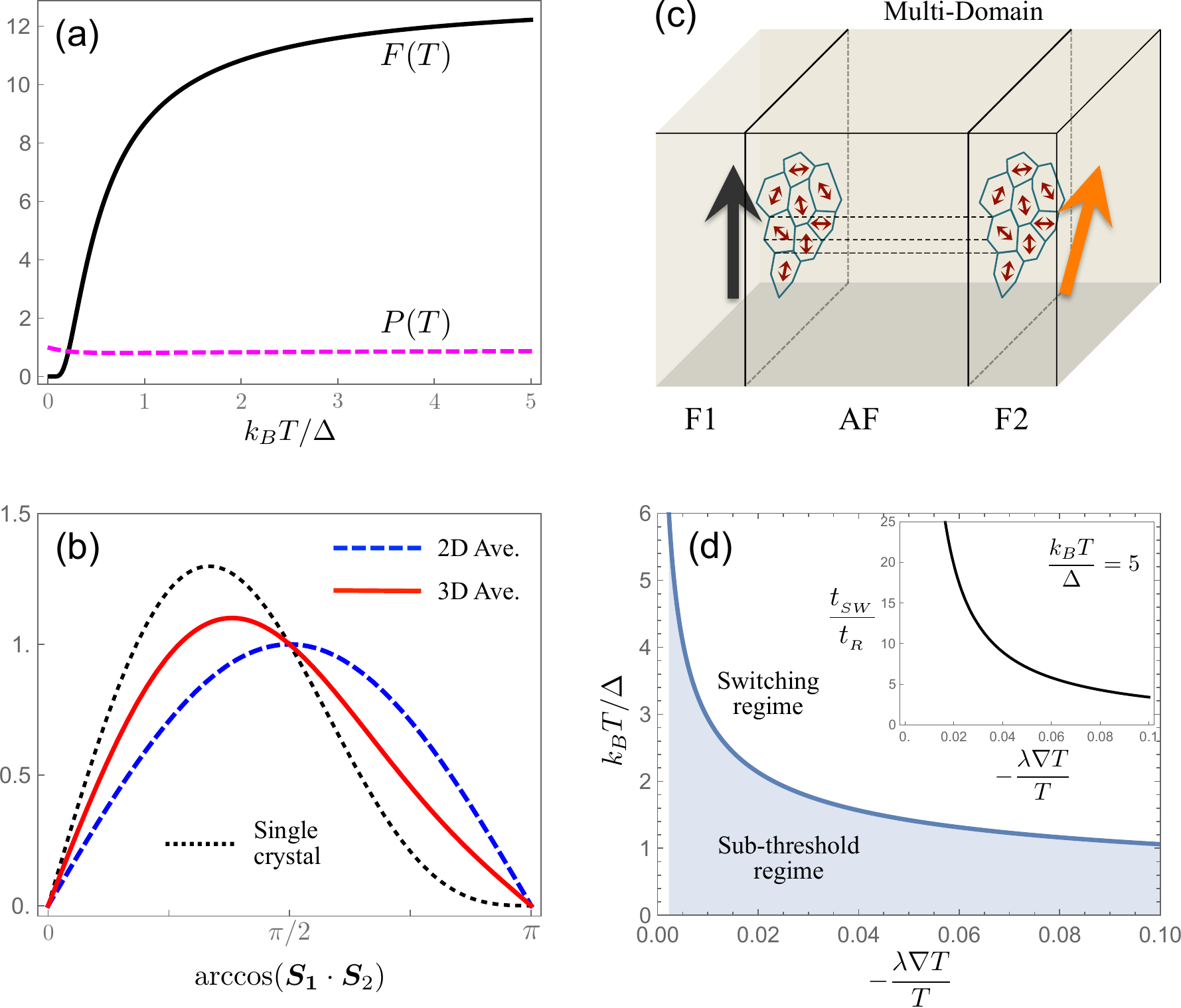}
	\caption{(a) $F(T)$ in Eq.~\eqref{eq:Ffunc} and the effective magnon spin polarization $P(T)$. (b) Angle dependence of the magnonic STT for single- and multi-domain AFs. (c) Domains are random in the lateral dimensions but they do not end along $x$. (d) Phase boundary separating the switching and the sub-threshold regimes for $N_F=10$, $\alpha/\pi=0.01$, $K_F/K_{AF}=0.1$, and $J_{AF}/J=100$. Inset: the switching time $t_{SW}$ (scaled by $t_R=2\pi\hbar/K_F$) as a function of the temperature gradient at $k_B T=5\Delta$.}
	\label{fig:switch}
\end{figure}

To linear order in $-\nabla T/T$, the deviation from local equilibrium $\delta f_\sigma^>\equiv f_\sigma^>-\bar{f}_\sigma$ satisfies
\begin{align}
 v_x^\sigma\frac{\partial\bar{f}_\sigma}{\hbar\partial\omega}(\hbar\omega_\sigma-\mu_\sigma)\frac{\partial_x T}{T}=\frac{\delta f_\sigma^>-\delta f_\sigma^<}{\tau_p} \label{eq:Boltzmann}
\end{align}
for $v_x>0$, and $\delta f_\sigma^<\equiv f_\sigma^<-\bar{f}_\sigma$ satisfies a similar Boltzmann equation for $v_x<0$. With those non-equilibrium distribution functions, we are able to calculate the spin current density in the AF (scaled in a number current density): $j_s=\int d^3k\left[ v_x^\uparrow (\delta f_{\uparrow}^>-\delta f_{\uparrow}^<)-v_x^\downarrow (\delta f_{\downarrow}^>-\delta f_{\downarrow}^<)\right]$, which, in general, can be solved only numerically. But to linear order in $J/k_BT\ll1$, we obtain a simple expression of the spin current density flowing into F2 as
\begin{align}
 j_s(L)=\frac{-\lambda\partial_x T}{T}\frac{4\pi J(k_BT)^2}{3\hbar^2v_s^2}F\left(\frac{k_BT}{\Delta}\right)(1+\cos\theta), \label{eq:js}
\end{align}
where $\lambda=v_s\tau_p$ is the effective momentum relaxation length. The function $F$ has the form
\begin{align}
 F(x)=\int_{1/x}^{\infty}dz\frac{\coth\frac{z}{2}-\frac{1}{z}}{2\sinh^2\frac{z}{2}}\left(z^2-\frac{1}{x^2}\right)^{3/2}, \label{eq:Ffunc}
\end{align}
which exhibits a saturation behavior at high temperatures as plotted in Fig.~\ref{fig:switch}(a). Regarding Eqs.~\eqref{eq:LLG} and~\eqref{eq:js}, we know that the net STT acting on F2 has an angle dependence $\sin\theta(1+\cos\theta)$ as plotted by the black dotted line in Fig.~\ref{fig:switch}(b). However, this angle function is problematic for magnetic switching. Since the Gilbert damping $\alpha\bm{S}_2\times\dot{\bm{S}}_2$ is proportional to $\sin\theta$, the STT will always lose at around $\theta=\pi$. Our STT acquires this special angle profile since we have modeled the AF as a single-domain system, in which the preset easy-axis breaks the rotational symmetry. Therefore, a remedy can be found by considering multi-domain average~\cite{note2}.

Suppose that the AF breaks into multi-domains in the transverse dimension with random N\'{e}el orders as depicted in Fig.~\ref{fig:switch}(c). So long as domains do not terminate in the thickness direction, the magnon transport solved earlier still holds within each individual domain. This situation allows us to average over all domains to obtain a magnonic STT that depends only on the relative angle $\theta$ between F1 and F2. After a straightforward calculation, we ends up with a (thermal) current-induced torque
\begin{align}
 \tau_s=j_mP(T)\bar{g}(\theta)\bm{S}_2\times(\bm{S}_1\times\bm{S}_2), \label{eq:final}
\end{align}
where $j_m=-\zeta\partial_xT/T$ is the magnon Seebeck current density (analogous to the charge current density of electrons) with $\zeta$ the Seebeck coefficient; and $P(T)$ is the effective spin polarization of magnon currents at temperature $T$. As plotted in Fig.~\ref{fig:switch}(a), $P(T)$ has a very weak temperature dependence and can be regarded as a constant (around $0.87$) at room temperature. In Eq.~\eqref{eq:final}, the angle profile $\bar{g}(\theta)$ depends on how spins randomly spread over different domains: (i) If the N\'{e}el vectors can point to any direction in space (a 3-D random), then $\bar{g}(\theta)=(2+\cos\theta)/2$. (ii) If the N\'{e}el vectors in different domains are restricted to be in-plane (a 2-D random), then $\bar{g}(\theta)$ will depend on both $\theta$ and the angle between $\bm{S}_2$ and the interface. However, when we consider a simple case in which $\bm{S}_2$ also rotates in the film plane, the angle profile reduces to $\bar{g}(\theta)=1$ so that the STT and the Gilbert damping share the same angle dependence. For both cases, the slop of $\tau_s$ at $\theta=\pi$ are now nonzero as shown in Fig.~\ref{fig:switch}(b), hence a finite threshold could be expected.

Since $\zeta$ is an increasing function of temperature while $P(T)$ is essentially a constant, the threshold temperature gradient decrease monotonically with an increasing temperature. Considering that $\Delta^2=Z^2J_{AF}K_{AF}$ and $v_s=ZaJ_{AF}/\hbar$ with $J_{AF}$ the Heisenberg exchange coupling of bulk spins, $K_{AF}$ the anisotropy constant and $Z$ the coordination number of the AF, we obtain the threshold temperature gradient (for the case of 3-D multi-domain average) as
\begin{align}
 \left( -\frac{\lambda\partial_xT}{T} \right)_{th}=\frac{3\alpha N_F}{\pi}\left( \frac{K_FJ_{AF}}{K_{AF}J} \right)\frac{1}{t^2F(t)}, \label{eq:threshold}
\end{align}
where $K_F$ is the anisotropy constant of F2, $N_F$ is the number of atomic layers of F2 in the thickness direction, and $t=k_BT/\Delta$ is the scaled temperature. From Eq.~\eqref{eq:threshold}, we see that a larger interfacial exchange coupling $J$ can lower the threshold temperature gradient; we can also reduce the threshold by raising temperature since $t^2F(t)$ in the denominator is an increasing function of $t$. In Fig.~\ref{fig:switch}(d), we draw the phase diagram of F2 versus $-\lambda\nabla T/T$ and $k_BT/\Delta$ using material parameters listed in the caption. The average switching time $t_{SW}$ is plotted in the inset at $k_BT=5\Delta$ (close to room temperature). If the effective magnon momentum relaxation length $\lambda$ is few nanometers, the threshold temperature gradient is evaluated to be somewhere between $0.1\sim1$ K/nm at room temperature using parameters of typical AF insulators such as MnF$_2$ and Cr$_2$O$_3$. Besides tunning temperature, the threshold value can be reduced by, for example, increasing the interfacial exchange coupling, reducing the intrinsic anisotropy of F2, \textit{etc.} Practically, to get a switching time on the order of nanoseconds (comparable to electronic switching), a temperature gradient of order $1$ K/nm shall be needed.

Since $j_s$ (or $\tau_s$) is odd in $\nabla T$, a reversed temperature gradient generates an inverse spin current. In this circumstance, $\bm{S}_2$ loses spin-up angular momenta, hence an opposite STT will drive $\bm{S}_2$ towards the antiparallel state. For the 3-D averaged case, the slop of $\tau_s$ has a larger absolute value at $\theta=0$ than at $\theta=\pi$ as shown in Fig.~\ref{fig:switch}(b). Therefore, the threshold for the reciprocal switching is relatively smaller---a vastly different behavior compared to the case of electrons~\cite{ref:STT,ref:Slon}. In fact, the actual difference in reciprocal switching is dominated by the exchange bias (EB) effect, which we have not considered in this Letter. While the EB on the F1/AF interface can stabilize the setup, the EB on the F2/AF interface will be a destructive effect. In order to minimize the influence of the latter, it is ideally to grow, field anneal, and train the F1/AF part first before growing the F2 layer. Alternatively, we can choose to operate the device slight above the blocking temperature.

The threshold in Eq.~\eqref{eq:threshold} is inversely proportional to $K_{AF}/J_{AF}$, which can be understood intuitively from a semiclassical perspective: In a spin-wave eigenmode of a collinear AF, the ratio of cone angles of the sublattice precessions $\theta_A/\theta_B$ is proportional to $(1+\sqrt{K_{AF}/J_{AF}})^2$~\cite{ref:AFMR,ref:AFSP}, thus the difference in their $z$-components $(\cos\theta_A-\cos\theta_B)$, \textit{i.e.} the net spin angular momentum, is proportional to $K_{AF}/J_{AF}$ under a small angle expansion. This fact suggests that the threshold will be lower if we find AFs with larger easy-axis anisotropy.

\textit{Discussion.}---Our proposed all-magnonics switching requires the coexistence of both spin-up and spin-down carriers, thus being unique to collinear AFs with easy-axis anisotropy. Different from electrons, thermally populated magnons bear spin polarizations along the easy-axis while transverse spins vanish.

Below we remark on the experimental feasibility and potential challenges to our prediction. It is difficulty to generate a sufficiently large temperature gradient using current-induced heating~\cite{ref:Igor}. But fortunately, with the development of heat-assisted magnetic recording, an extremely large temperature gradient of $10$ K/nm has been achieved by optically excited near-field transducer~\cite{ref:HAMR}; our predicted threshold less than $1$ K/nm is well within its capacity. On the other hand, momentarily laser heating introduces side effects such as lattice distortion, which can break the intrinsic uniaxial anisotropy through magnetostriction. On top of this, structural imperfections such as interfacial lattice mismatch may lead to additional anisotropies. Nonetheless, as long as those induced effects do not destroy the collinear N\'{e}el ground state, they all appear to be perturbative that only deform the circular polarization into an elliptical one. Although the magnitude of magnon spin is reduced when the polarization becomes elliptical, the coexistence of both spin species persists.

A possible way of detection is to attach a heavy metal layer adjacent to the free F and monitor the spin Seebeck effect via the inverse spin Hall voltage~\cite{ref:SSE}. If the free F switches, the inverse spin Hall voltage also flips sign. The choice of materials is broad since the major requirement is that the AF has a single easy-axis. For example, MnF$_2$, FeF$_2$, and Cr$_2$O$_3$ are all legitimate candidates. There is no special requirement on the choice of F, but a potential complication is whether it can couple to the AF with minimal lattice distortion. If both F1 and F2 are insulating YIG, the fabrication might be difficult; if one of the Fs is metallic such as permalloy, then interfacial magnon-electron scattering~\cite{ref:Duine,ref:Slons} must be taken into account. Since we intend to draw a general physical picture and a proof-of-concept of the magnonic STT, it goes beyond the scope of this Letter to nail down specific materials.

Our model is only valid for $L\gg\lambda$, which becomes increasingly inaccurate when the film thickness is getting smaller. For ultrathin AF films in which both momentum and spin are ballistic, the Boltzmann transport picture breaks down. In this regime, the wave vector $k_x$ is quantized due to geometrical confinement, and the interlayer physics manifests as a magnon-mediated RKKY interaction~\cite{ref:Ran}.

After submitting our manuscript, we became aware of a recent paper on the same topic~\cite{ref:MMT}, where the modeling is very different. In this Letter, magnons reside in the AF and acquire polarization from exchange splitting in the vicinity of interfaces. So when F1 is spin-up, the majority carriers are spin-up. By sharp contrast, Ref.~\cite{ref:MMT} assumes that magnons are generated in F1 and then injected into the AF~\cite{ref:KC} under thermal gradient, thus the majority carriers are spin-down. In our opinion, in order to switch F2 from spin-down to spin-up, the majority carriers should be spin-up. The difference between Ref.~\cite{ref:MMT} and our result arouses an interesting issue to be settled by experiments.

\begin{acknowledgments}
  R.C.\ acknowledges insightful discussions with Weiwei lin, Chia-Ling Chien, Dazhi Hou, and Shulei Zhang. R.C.\ and D.X.\ were supported by the Department of Energy, Basic Energy Sciences, Grant No.~DE-SC0012509.
\end{acknowledgments}


\begin{thebibliography}{20}
\bibitem{note2} Domain boundaries not necessarily coincide with grain boundaries. Multi-domain structures also prevent the formation of exchange spring.
\end{thebibliography}
\end{document}